\documentclass{aa} 

\newcommand\nice[1]{#1}    \newcommand\subm[1]{}   

\nice{\newcommand\dosingle[1]{#1}  \newcommand\dodouble[1]{ }} 
\subm{\newcommand\dosingle[1]{ }   \newcommand\dodouble[1]{#1}} 

\dodouble{\documentclass[referee]{aa}}

\nice{ \usepackage{color} }
 \newcommand\recenteditsone[1]{ { #1} }

\newcommand\bfrevisedversion[1]{ {#1} } \newcommand\bfrevisedversionstart{ } \newcommand\bfrevisedversionstop{ }

\usepackage{graphicx}  

\usepackage{amsfonts}
\usepackage{mathpi}

 \usepackage{hyperref}
\nice{ 
\providecommand{\eprint}[1]{\href{http://arxiv.org/abs/#1}{[arXiv:#1]}}

\providecommand{\url}[1]{\href{#1}{#1}}

}

\providecommand{\adsurl}[1]{} 
\def\SSS{Sect.~}

\nice{ \newcommand\mycaptionfont{\protect\footnotesize} }

\def\gtapprox{\,\lower.6ex\hbox{$\buildrel >\over \sim$} \, }
\def\ltapprox{\,\lower.6ex\hbox{$\buildrel <\over \sim$} \, }
\def\propapprox{\,\lower.6ex\hbox{$\buildrel \propto\over \sim$} \, }

\def\arcs{\ifmmode {'' }\else $'' $\fi}     
\def\arcm{\ifmmode {' }\else $' $\fi}       

\def\fr7{7$ \hskip -0.9ex \vrule height0.8ex width0.8ex depth-0.73ex
                                                     \hskip0.1ex$}
\def\frtoday{Le\space\number\day\space\ifcase\month\or
  janvier\or f\'evrier\or mars\or avril\or mai\or juin\or
  juillet\or ao\^ut\or septembre\or octobre\or novembre\or 
d\'ecembre\fi\space \number\year}

\newcommand\cqg{ClassQuantGra}   %
\newcommand\BASI{Bull. Astr. Soc. India}

\newcommand\hMpc{\mbox{$h^{-1}$ Mpc}}
\newcommand\hGpc{\mbox{$h^{-1}$ Gpc}}

\newcommand\Omm{\Omega_{\mbox{\rm \small m}}}
\newcommand\Omb{\Omega_{\mbox{\rm \small b}}}
\newcommand\Omtot{\Omega_{\mbox{\rm \small tot}}}

\newcommand\cGW{c_{\mbox{\rm \tiny GW}}}
\newcommand\cST{c_{\mbox{\rm \tiny ST}}}
\newcommand\cGR{c_{\mbox{\rm \tiny GR}}}
\newcommand\cEM{c_{\mbox{\rm \tiny EM}}}

\newcommand\deff{d_{\mbox{\rm eff}}}

\newcommand\ddotrtopo{\ddot{\bf{r}}_{\mbox{\rm topo}}}

\title{A weak {acceleration effect} 
due to residual gravity in a multiply connected universe}

\author{Boudewijn F. Roukema \inst{1}
\and Stanislaw Bajtlik \inst{2}
\and Marek Biesiada \inst{3}
\and Agnieszka Szaniewska \inst{1}
\and Helena Jurkiewicz \inst{1}
}

\institute{Toru\'n Centre for Astronomy, N. Copernicus University,
ul. Gagarina 11, PL-87-100 Toru\'n, Poland
\and
Copernicus Astronomy Centre, ul. Bartycka 18, PL-00-716 Warsaw, Poland
\and
Astronomy \& Cosmology Department, University of Silesia, Uniwersytecka 4, 
PL-40-007 Katowice, Poland
}

\date{\frtoday}

\titlerunning{Weak $\Omega_{\mbox{X}}$ in multiply connected universe}
\authorrunning{Roukema et al.}

\abstract
{{{
Understanding dark energy and measuring the topology of the Universe
are two of the biggest open questions in physical cosmology.
It was previously shown that multiple connectedness, 
via the twin paradox of special 
relativity, provides a novel physical justification for an
assumption of the standard FLRW model: 
it implies a favoured space-time splitting (comoving coordinates).
}}}
{Could cosmic topology also imply dark energy?}
{We use a weak field (Newtonian) approximation of gravity and 
consider the gravitational effect from distant, multiple copies of 
a large, collapsed (virialised) object today (i.e. a massive galaxy cluster), 
taking into account
the finite propagation speed of gravity, in a flat, multiply connected universe,
and assume that due to a prior epoch of fast expansion (e.g. inflation), the
gravitational effect of the distant copies is felt locally, 
from beyond the na\"{\i}vely calculated horizon.}
{
{{We find that for a universe with 
a $\mathbb{T}^1 \times \mathbb{R}^2$ spatial section,
the residual Newtonian gravitational force (to first order)
provides an anisotropic effect that {{\em repels}} test
particles from the cluster in the compact direction, 
in a way algebraically similar to that of 
dark energy.
For a typical test object at comoving 
distance $\chi$ from the nearest dense
nodes of the cosmic web of density perturbations,
the pressure-to-density ratio 
$w$ of the equation of state in an FLRW universe, 
is $w \sim - (\chi/L)^3$, where 
$L$ is the size of the fundamental domain, i.e. of the
Universe. Clearly, $|w| \ll 1$. For a $\mathbb{T}^3$ spatial section of exactly 
equal fundamental
lengths, the effect cancels to zero. 
For a $\mathbb{T}^3$ spatial section of unequal fundamental lengths, the 
acceleration effect
is anisotropic in the sense that it will {{\em tend to equalise the three
fundamental lengths.}}
}}}
{
Provided that at least a modest amount of inflation occurred in the early
Universe, and given some other conditions, multiple connectedness
 {{\em does}} generate an effect similar to that of dark energy, but
the amplitude of the effect at the present epoch is too small 
to explain the observed dark energy density 
{{and its anisotropy makes
it an unrealistic candidate for the observed dark energy.}}
}


\begin{document}

\maketitle

\dodouble{\clearpage} 


\newcommand\fpotentials{
\begin{figure}[ht]
\centering
\includegraphics[width=9cm]{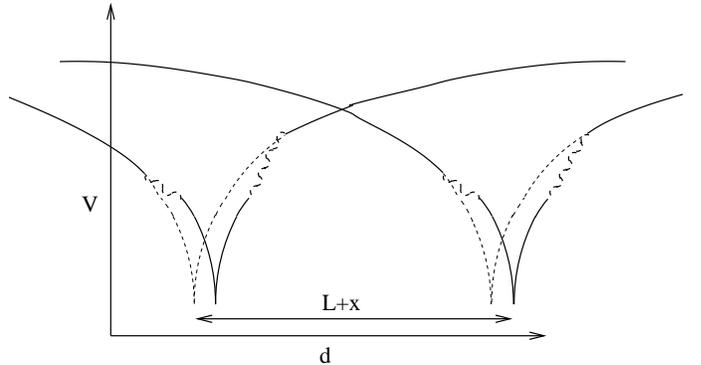}
\caption[distantly generated potentials]{ 
\nice{ \mycaptionfont }
{Schematic diagram showing the potential to a single, massive
object in a multiply connected universe of comoving size $L$, which is perturbed
to the right (of the figure) by a small physical distance $x$. The information 
about the changes in the potential generated by this object travels outward from the
object at a finite speed, $\cGW$. If $L$ is large (about a Hubble length in size), 
then at or close to the position of the object, the 
potential due to the (distant) topological image remains that of the 
unperturbed topological image, until a Hubble time has passed. 
Anthropomorphically, we could say that during about a Hubble time following
the perturbing event,
the object ``does not believe'' that its topological image has been perturbed.
Solid thick curves show the perturbed potentials, dashed thick curves show 
the original potentials, wavy thick curves symbolically show the information 
about the changed position of the
object being transmitted by gravitational waves.
}
\label{f-potentials}
}
\end{figure}
} 

\newcommand\fflatxL{
\begin{figure}[ht]
\centering
\nice{ \includegraphics[width=8cm]{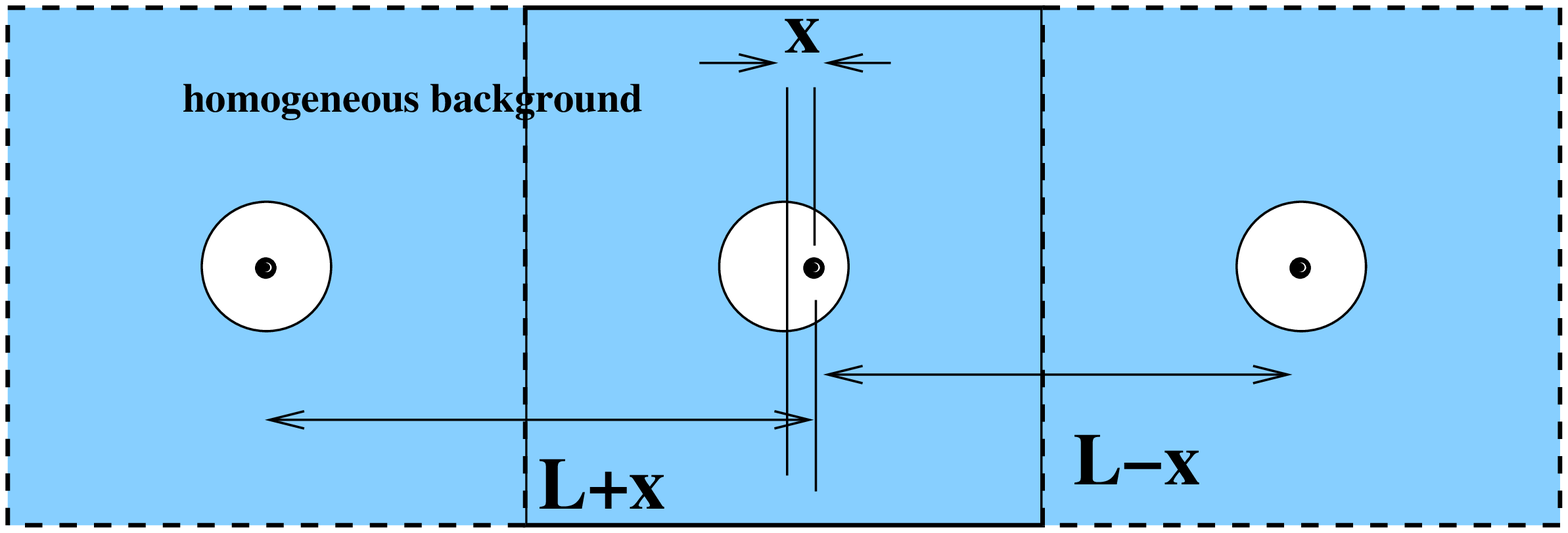} }
\subm{ \includegraphics[width=12cm]{topograv_flat} }
\caption[massive cluster self-gravity]{ 
\nice{ \mycaptionfont }
     {\bfrevisedversion{Self-gravity of a large cluster.} A 
       flat, toroidal Universe model of comoving side
       length $L$, filled with what is assumed to be a homogeneous
       density field except for one massive, collapsed object
       (e.g. cluster of galaxies) shown as a black spot surrounded by an
       empty sphere from which the matter forming it was originally
       distributed. This object is slightly perturbed, by physical 
       distance $x$, from its original position towards one of its
       adjacent virtual copies in the comoving (``apparent'')
       space. Since the potential in the ``central'' copy of the
       fundamental domain is determined by the two adjacent copies of the
       fundamental domain, the object ``perceives'' the adjacent
       topological images in their original positions.}  
\label{f-flatxL}
}
\end{figure}
} 

\newcommand\fflatxLtwo{
\begin{figure}[ht]
\centering
\nice{ \includegraphics[width=8cm]{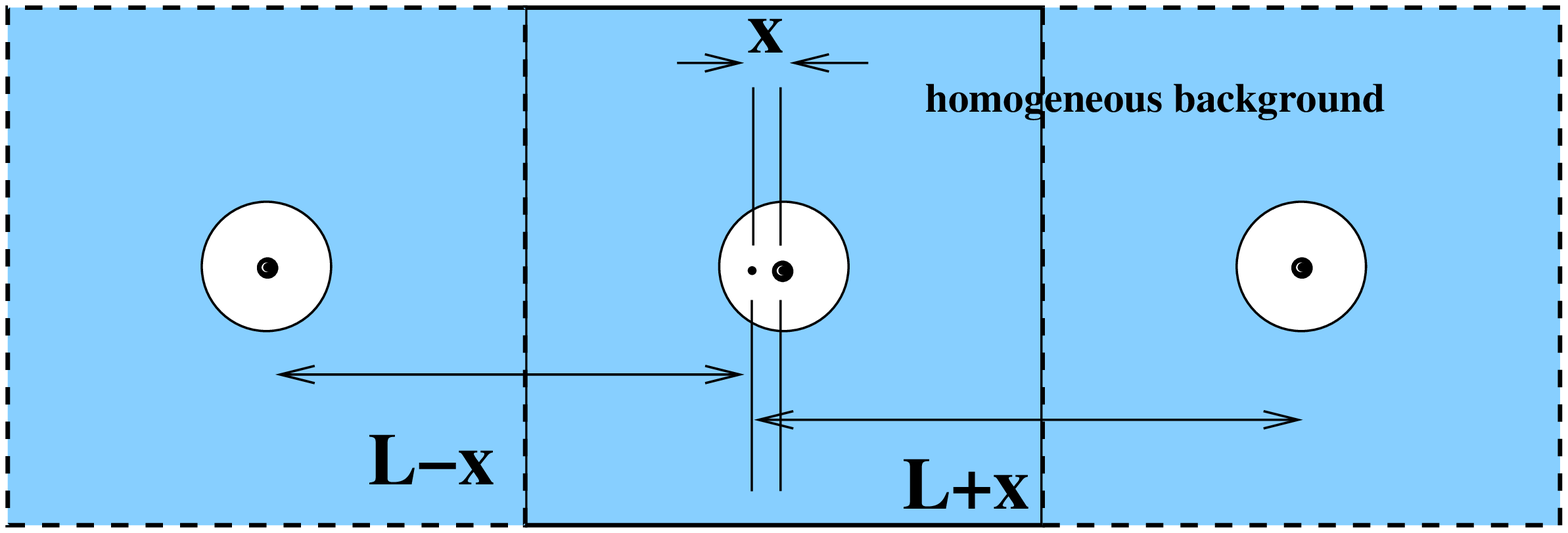} }
\subm{ \includegraphics[width=12cm]{topograv_flat2} }
\caption[small test object near massive cluster]{ 
\nice{ \mycaptionfont }
{\bfrevisedversion{Effect relative to the cosmic web.}
As in Fig.~\protect\ref{f-flatxL}, except
that the big black spot representing a massive, collapsed object is no
longer perturbed, and instead, a small test object located physical distance
$x$ from the cluster, along the line separating the two clusters, is shown. }
\label{f-flatxLtwo}
}
\end{figure}
} 

\newcommand\fflatxLthD{
\begin{figure}[ht]
\centering
\nice{ \includegraphics[width=8cm]{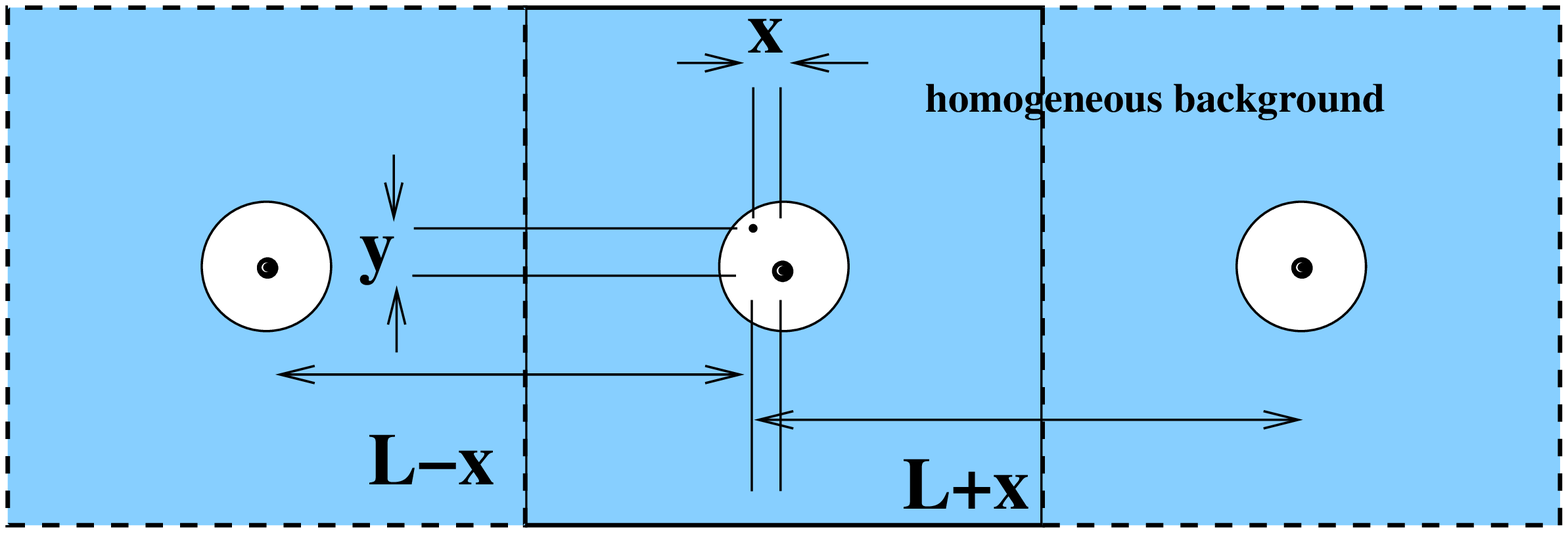} }
\subm{ \includegraphics[width=12cm]{topograv_flat3d} }
\caption[small test object near massive cluster --- 3d]{ 
\nice{ \mycaptionfont }
{{{
\bfrevisedversion{Effect relative to the cosmic web: }
three-dimensional version of Fig.~\protect\ref{f-flatxLtwo}. 
The small test object is separated from the 
cluster by $x$ along the direction of the generator $g_x$,
and by $y$ and $z$ in orthogonal directions. The $z$ direction is not
shown. Topological images of the cluster at $i=\pm1, j=k=0$ are shown.
}
}}
\label{f-flatxLthD}
}
\end{figure}
} 

\newcommand\ftwotorusmod{
\begin{figure}[ht]
\centering
\nice{ \includegraphics[width=8cm]{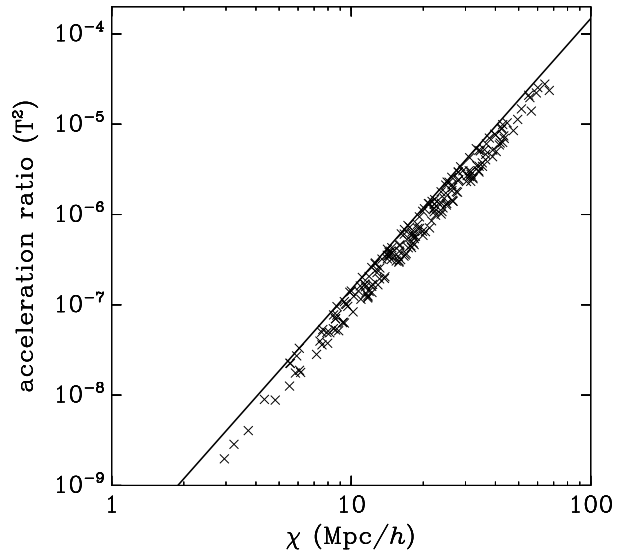} }
\subm{ \includegraphics[width=12cm]{fig5a} }
\caption{ 
\nice{ \mycaptionfont }
{
\bfrevisedversion{Ratio of the topological acceleration effect 
(the effect relative to the cosmic web)
over the normal gravitational 
infall, shown as a function of the distance of a test point to the cluster
exerting the greatest gravitational pull on it, using a Hubble Volume
simulation by the Virgo Supercomputing Consortium 
(see \SSS\protect\ref{s-numerical-method}).
This plot shows the {$\mathbb{T}^1$} case showing the ratios of the modulus
of the accelerations.
The line shows the analytical estimate $4 \left(\frac{\chi}{L}\right)^3$
[see Eqs~(\protect\ref{e-test-object-threed-xyz}) 
and (\protect\ref{e-w-estimate})].}
}
\label{f-twotorusmod}
}
\end{figure}
} 

\newcommand\ftwotorusx{
\begin{figure}[ht]
\centering
\nice{ \includegraphics[width=8cm]{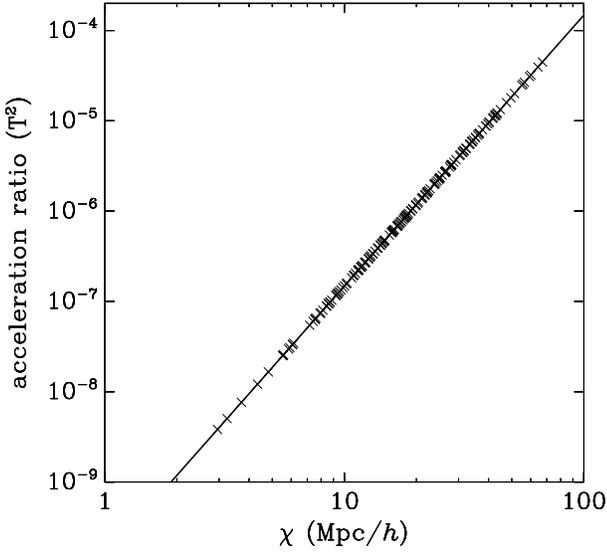} }
\subm{ \includegraphics[width=12cm]{fig5b} }
\caption{ 
\nice{ \mycaptionfont }
\bfrevisedversion{{{As per Fig.~\protect\ref{f-twotorusmod}, 
showing the (negative of the) topological-to-normal ratio
of the $x$ axis accelerations in the $\mathbb{T}^1$ case
and the function $4 \left(\frac{\chi}{L}\right)^3$. All test points were found
(as expected)
to have opposite signs in the $x$ direction topological and normal accelerations;
hence, the negative was used.
}
}}
\label{f-twotorusx}
}
\end{figure}
} 

\newcommand\fthreetorusmod{
\begin{figure}[ht]
\centering
\nice{ \includegraphics[width=8cm]{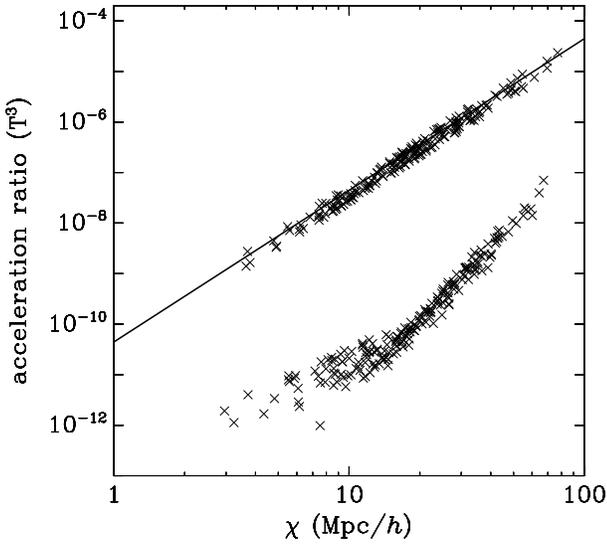} }
\subm{ \includegraphics[width=12cm]{fig6a} }
\caption{ 
\nice{ \mycaptionfont }
\bfrevisedversion{{{As per Fig.~\protect\ref{f-twotorusmod}, 
showing the ratios of the modulus
of the accelerations for the $\mathbb{T}^3$ case.
The lower scattering of points show the case with three equal fundamental
lengths and the upper scattering of points shows the case with 
$\delta_e=\delta_u = 0.1$.
The line shows the analytical estimate $12 \delta_e \left(\frac{\chi}{L}\right)^3$
[see Eqs~(\protect\ref{e-test-object-threed-unequal-shortx}) 
and (\protect\ref{e-w-estimate})].
}
}}
\label{f-threetorusmod}
}
\end{figure}
} 

\newcommand\fthreetorusx{
\begin{figure}[ht]
\centering
\nice{ \includegraphics[width=8cm]{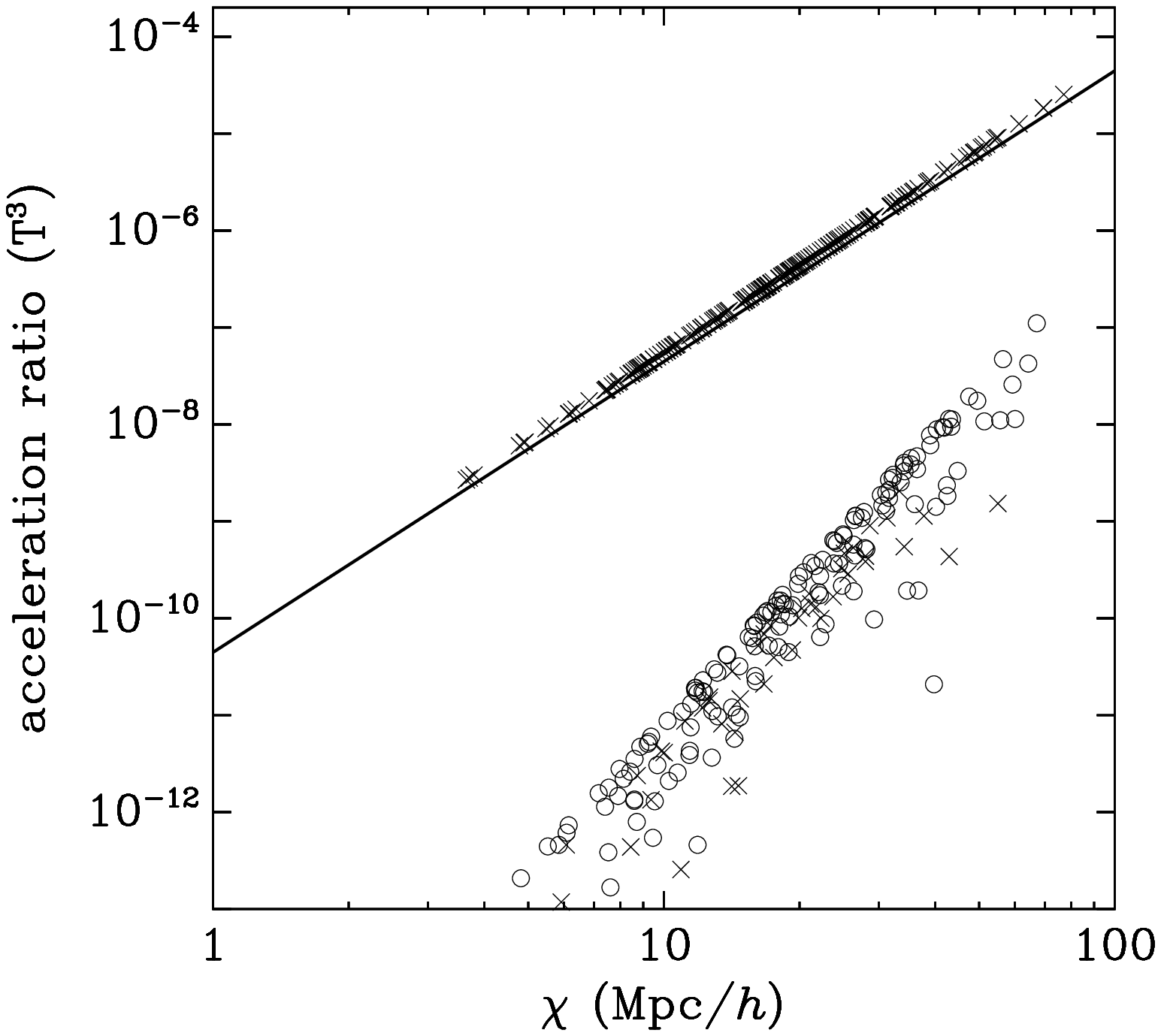} }
\subm{ \includegraphics[width=12cm]{fig6b} }
\caption{ 
\nice{ \mycaptionfont }
\bfrevisedversion{{{As per Fig.~\protect\ref{f-threetorusmod}, 
showing the (negative of the) 
ratios of the $x$ axis accelerations in the $\mathbb{T}^3$ case with
equal fundamental lengths (lower ``x'' points) and unequal lengths 
(upper ``x'' points).  Positives of the ratios, i.e. when a test
point is {\em attracted} by the topological residual force in the same
direction as the local gravitational force, are shown as circles. This latter
situation
only occurs for the case of equal fundamental lengths, where the topological
force cancels perfectly according to the analytical calculation.
The line shows $12\delta_e \left(\frac{\chi}{L}\right)^3$.
}
}}
\label{f-threetorusx}
}
\end{figure}
} 


\section{Introduction}

Two of the biggest open questions in physical cosmology are (1) the
interpretation of the cosmological constant (or some other form of dark energy),
and (2) the global shape of the Universe, including both curvature and
topology. The former is empirically known to exist from many
different, more or less independent, observational approaches
{(the cosmic concordance model, with matter density
$\Omm \approx 0.3,$ cosmological constant/dark energy $\Omega_\Lambda
\approx 0.7$, Hubble constant $H_0 \approx 70$~km/s/Mpc, baryon
density $\Omb \approx 0.05$, \nocite{LahavLiddle04}(e.g.  {Lahav} \& {Liddle} 2004) 
)}, and observations have also suggested
an answer to the latter, but the evidence is far from being
conclusive, and, in principle, a final answer may be totally
beyond the reach of any observations.

The analysis of empirical data that suggests the shape of the Universe 
consists of recent analyses 
of the cosmic microwave background observations by the
WMAP satellite. These analyses
found the temperature fluctuation map 
to be better modelled by a multiply-connected
model of the Universe, for a Poincar\'e dodecahedral space (PDS) as the
3-manifold of comoving space, rather than by an ``infinite'' flat space
 \nocite{LumNat03,RLCMB04,Aurich2005a,Aurich2005b,Gundermann2005}(e.g.  {Luminet} {et~al.} 2003; {Roukema} {et~al.} 2004; {Aurich} {et~al.} 2005a, 2005b; {Gundermann} 2005).

These analyses may or may not be supported by future observations.
One of the most independent tests of the PDS hypotheses will be high precision
estimates of the total density parameter $\Omtot$. The hypotheses
of these authors geometrically require $\Omtot$ to be strictly
greater than unity and $\Omtot$ should lie in the 
predicted range $1.01 \ltapprox \Omtot \ltapprox 1.02$ 
(different authors using different methods make slightly different 
predictions). Future observations which yield, for example, 
$\Omtot = 1.001 \pm 0.001$, would rule out these hypotheses to a significance
level of $9\sigma$.

Independently of the present hypotheses, 
the question of the {\em physical} consequences 
of cosmic topology will remain. 

We know that {\em locally}, geometry and density are directly related, via
general relativity. But we have very few hints as to how {\em global}
geometry should relate to other physical properties. General relativity
is a local theory, not a global one.

One known effect is that cosmic topology defines the comoving 
reference frame.

It was shown by \nocite{UzanTwins02}{Uzan} {et~al.} (2002) and \nocite{LevinTwins01}{Barrow} \& {Levin} (2001) 
that resolving the twin paradox of special 
relativity in a multiply-connected universe is different than in
simply connected Minkowski space. Moreover, multiple connectedness implies
a favoured space-time splitting. The authors show that this 
must correspond to the comoving reference frame. 

{In other words,}
this provides a global geometrical motivation for Weyl's postulate,
i.e. the postulate that ``the world lines of the galaxies form a
three-bundle of non-intersecting geodesics orthonormal to a series of
space-like hypersurfaces''.


Could it be possible that cosmic topology could also provide 
a simple explanation for the 
{observed density of dark energy}?

If there is such a mechanism, then this would most naturally be
able to explain the fact why 
{dark energy
starts to dominate} at
the present epoch if this mechanism were linked to a property of 
the present cosmological epoch. 

One property of the present epoch is that density perturbations have
collapsed due to gravity and formed virialised structures, of which
the most dense at the present epoch are galaxy clusters, the largest
which typically are of mass
$M \sim 10^{15} M_{\odot}$, which are formed from most of the material
within a region of linear comoving size $\sim 10-20$~Mpc.

What property of virialised structures is different between the
simply connected case and the multiply connected case?

The difference is that {\em in the covering space (apparent space)},
the spatial distribution of distant objects is homogeneous and uncorrelated
with the local distribution if space is simply connected, while 
if space is multiply connected, then the spatial distribution of distant
objects in the covering space is {\em not} random. 

{\em Given the non-random distribution of multiple topological 
images of objects such
as clusters in the covering space, and the possibility
that there was causal contact between them at some time in the past 
(e.g. due to some moderate amount of inflation), is it possible that
the residual gravitational effect due to distant, multiple topological images 
of a single object yields a gravitational effect 
different from that normally expected from the assumption of large-scale
unperturbed, homogeneity?}

In \SSS\ref{s-method}, 
the assumptions made in 
considering this gravitational effect, in the covering space, 
between multiple copies of a
``small'', massive object, perturbed from its initial position, 
are listed.

In \SSS\ref{s-results}, the effect is calculated. 

In \SSS\ref{s-discuss}, the results of this calculation are discussed,
and in \SSS\ref{s-conclu} we conclude.

For a short, concise review of the terminology, geometry and 
relativistic context of cosmic topology, see \nocite{Rouk00BASI}{Roukema} (2000) 
(this is now slightly outdated, but is sufficient for beginners).
For in-depth review papers  see, 
e.g. \nocite{LaLu95,Lum98,Stark98,LR99}{Lachi\`eze-Rey} \& {Luminet} (1995); {Luminet} (1998); {Starkman} (1998); {Luminet} \& {Roukema} (1999); workshop proceedings 
are in \nocite{Stark98}{Starkman} (1998) and following articles,
and \nocite{BR99}{Blanl{\oe}il} \& {Roukema} (2000). 
For comparison and classification of different {\em observational} strategies,
see e.g. \nocite{ULL99b,LR99,Rouk02topclass,RG04}{Uzan} {et~al.} (1999); {Luminet} \& {Roukema} (1999); {Roukema} (2002); {Rebou\c{c}as} \& {Gomero} (2004). 
What might
be considered as one of the most striking {\em theoretical} results,
in the sense of providing 
a direct link between the FLRW model and multiple-connectedness,
is the implication of a favoured reference frame,
which must coincide with the comoving reference 
frame: see \nocite{LevinTwins01}{Barrow} \& {Levin} (2001) and \nocite{UzanTwins02}{Uzan} {et~al.} (2002). 

{Note that some previous work has been done on
  possible links between topology and dark energy, via the Casimir
  effect. See e.g. \nocite{Nouri05}{Ahmadi} \& {Nouri-Zonoz} (2005) and references
  therein. \nocite{LachiezeReyCTP99}{Lachi{\`e}ze-Rey} (1999) estimated that for a $\mathbb{T}^1$ topology, 
  the effect would be at least 50 orders of magnitude too weak.}

\section{Assumptions and calculational choices}
\label{s-method}

Initially, 
in \SSS\ref{s-cluster-self-gravity},
we consider the {\em self}-gravity of a cluster to itself,
provided that the cluster has been perturbed from its initial, comoving
position. 
In \SSS\ref{s-test-object},
the more 
realistic case of essentially stationary (in comoving coordinates)
clusters distributed in the nodes of the cosmic web and the effect that
their multiple images have on small ``test objects'' nearby, is 
considered.

\subsection{Self-gravity of a large cluster}
\label{s-cluster-self-gravity}

\fpotentials

The following assumptions and choices are made:
\begin{list}{(\arabic{enumi})}{\usecounter{enumi}}
\item
 Newtonian approximation of gravity
\item
 a flat covering space, $R^3$
\item 
 calculations are made in the covering space 
\item 
 a region at least a few times the size of the injectivity diameter [assumed
to be on a size scale approximately similar to that of the 
diameter of the surface of last 
scattering (SLS)]
has been in causal contact (e.g. due to inflation) 
\item 
the gravitational potential induced from density perturbations at 
large, supra-SLS distance scales mostly cancels out due to homogeneity, but
since multiple topological images are non-randomly distributed, their total
contributions to the potential may not fully cancel and must be calculated
explicitly
\item 
up to a few tens of Mpc around a big cluster, the only non-uniform,
long-distance contributions to the local potential are from its
multiple topological images
\item 
{\em time lapse assumption}: 
the potential from these distant images corresponds to the state of 
these images as they were at cosmological time $t_1$,  
about a Hubble time in the past 
($t_1 < t_{\mbox{\rm \small recombination}} \ll t_0$); 
at that cosmological time (possibly pre-inflation, see (4) above), 
these images consisted of  
not-yet-collapsed density
perturbations, which had not yet had time to be significantly displaced towards
other dense regions --- both this assumption and the following (8) 
make the physically standard assumption that the 
speed of transmission of gravitational waves $\cGW$ is equal to that of the
special and general relativistic space-time constants and the
electromagnetic transmission speed: $\cGW = \cST = \cGR = \cEM$; see 
\nocite{EllisUzanC03}{Ellis} \& {Uzan} (2005).
\item 
{\em time lapse corollary}: 
if the cluster is spatially offset (perturbed) from its original 
position, then the long-range contributions to the gravitational potential
are only felt locally after at least a Hubble time: on a time scale much 
less than a Hubble time, {\em local} calculations can validly assume that
only the cluster moves, not its topological images, since they have not
yet received the information that the topological images have moved 
(see Fig.~\ref{f-potentials}).
\end{list}

(1), (2): Assumptions (1), (2) are identical to those made for most 
cosmological $N$-body
simulations of galaxy formation in the FLRW model,
e.g. \nocite{RPQR97}{Roukema} {et~al.} (1997), \nocite{Bagla05}{Bagla} (2005) and references therein. 
In other words, the ``volume effect'' and the 
``backreaction effect'' are considered negligible - see 
\nocite{BuchCarf03}{Buchert} \& {Carfora} (2003) for the basic equations.
(The volume and backreaction effects correct for 
the fact that the Universe does not have an exactly homogeneous FLRW metric, 
it is only {\em approximately} FLRW.) 

(3): Particle-Mesh (PM) 
cosmological $N$-body codes of galaxy formation 
(and also those which combine PM on a large scale with alternatives
on smaller scales) almost always assume a spatial 3-manifold which is 
the three-torus, $\mathbb{T}^3$, so calculations are made in the {fundamental 
domain} --- there is no need to use the covering space, 
so point (3) is not needed for these type of $N$-body codes.

Most direct $N$-body codes and tree codes (TC) also assume a 3-torus model,
but make calculations in the covering space rather than in the fundamental
domain. 
This is the choice (3) made here: 
calculations in the covering space are (usually) 
geometrically simpler in the covering
space than in the fundamental domain. 

(4): If there is no causal contact beyond the SLS, then no effect from
the topologically lensed images can occur. Here, the
case that the causal radius is much greater than the radius of the
SLS, e.g. due to an earlier, moderate amount of inflation, is
considered. \nocite{LindeTopo04}{Linde} (2004) has recently argued that for zero or negative
curvature, multiply-connected universes are {\em more} likely than 
simply-connected universes, and that these multiply-connected universes
are expected to have undergone a moderate amount of inflation.

(5), (6): While the assumption that most large-distance effects on the 
potential should approximately cancel each other 
is likely to be a good approximation, the possible 
{\em non}-cancelling of the potential
due to large-distance multiple images of a single cluster is likely to be
of a similar order of magnitude to the effects which we are assuming to 
cancel under assumptions (5) and (6). 

Nevertheless, we are interested in investigating whether {\em any} 
long-distance gravitational 
effect, in addition to contributions from local inhomogeneities, 
occurs due to topological imaging. 

If the result were a large effect,
then we would have to verify that it is fully self-consistent with the effects
from ``randomly'' distributed objects.

To some degree, we could expect that the effect has already been partially
modelled in PM and TC cosmological $N$-body simulations, without making
the assumptions (5) and (6). However, these simulations generally make 
``realistic'' assumptions for long-distance gravitational effects, which
mean some combination of assuming long-distance homogeneity and 
assuming an infinite speed $\cGW$ of the transmission of gravity, i.e. 
ignoring assumptions (7) and (8).

(7): The time lapse between transmission of information from a gravitational
source (density perturbation) and its arrival at a ``target'' point in 
comoving space is normally ignored in $N$-body simulations of galaxy formation:
gravity is assumed to be transmitted instantaneously. This is usually a
reasonable approximation, since the gravitational effect from distance sources
can generally be approximated (e.g. as in the top-down tree code simulations)
by considering the mass in a large, distant cube of space, 
which subtends a small angle at the ``target'' point, as a single, very massive
point object.

However, 
for topological gravity effects, we {\em do} take into account this time lapse.

(8): The initial density perturbations from which a cluster formed are
``not aware'' of the fact that the cluster later on collapses
gravitationally and moves towards a neighbouring potential well.

Assumption (8) gives us the first of the two 
effects which we will calculate. 

First consider a static cluster, which does not move. This is separated
from one of its ``adjacent'' topological images by a ``generator'' $g_x$. 
In general,
this is an isomorphism in the covering space. In the simplest 3-torus case, 
it is a translation, and can be thought of as a vector in Euclidean 3-space 
(the covering space). It is also separated from one of its adjacent topological
images in the opposite direction, i.e. by the generator $-g_x$.

\fflatxL

Now consider 
a cluster which has {\em moved} a small distance from its ``initial'' location 
in comoving 
space a short time ago. The contributions to the gravitational potential,
near to this cluster, from the other topological images --- which are 
distant --- remain the same as they were {\em before} the perturbation occurred.
This is shown in Fig.~\ref{f-potentials}.

\recenteditsone{
If we consider the potentials close to the object to be the potentials 
relevant for making local calculations about acceleration (or worldlines),
then similarly to the way luminosity distance $d_L$ is defined in terms of 
the observational flux $f$ and intrinsic luminosity $L$, 
$f = L/ (4\pi d_L^2)$, 
we can define the {\em effective comoving distance} $\deff$ 
(for a closer analogy with luminosity distance, we
could also call this the ``gravity distance'') in terms of the intrinsic mass 
$m$ and the component of locally felt acceleration $\ddot{x}$ due to the 
distant object 
(the ``observed'' gravitational acceleration), i.e. $\deff$ is the distance
satisfying
\begin{equation}
\ddot{x} =  - \frac{Gm}{\deff^2}.
\label{e-defn-deff}
\end{equation}
In other words, $\deff$ is {\em the comoving 
distance to a distant object implied by the local shape of its potential,
taking into account the standard value of 
$\cGW \approx 3\times 10^8m/s$ rather than the na\"{\i}ve Newtonian approximation 
$\cGW = \infty$}.

Using this concept of distance, 
a cluster which has {\em moved} from its ``initial'' location 
in comoving 
space is nearer to one (or two or three, depending on the direction of 
motion) of its topological images, since in standard physics, $\cGW$ is finite.
} 

Hereafter, we use the {\em effective comoving distance} unless otherwise
stated.

Let us consider the component of its motion 
towards one of its three adjacent images, so that the cluster is displaced 
by distance $x$ (in ``physical'' coordinate units) 
from its ``initial'' location in the direction of
generator $g_x$ of comoving length $L$, as shown in Fig.~\ref{f-flatxL}. 
Since we assume Newtonian gravity
at the present epoch, $L$ is also the relevant distance in proper units.

{\em Given points (1) to (8), we now have a cluster which feels unequal 
gravitational pulls from a pair of its closest topological images: it is 
slightly less than $L$ from one image, and slightly more than $L$ from 
the opposite image. The net result should be a gravitational pull towards
the former.}

This self-gravity effect should be absent in any $N$-body simulation which
assumes gravity is transmitted instantaneously, since the effects of the
two adjacent topological images will always perfectly cancel if 
$\cGW = \infty$ is assumed.

\subsection{Effect relative to the cosmic web}
\label{s-test-object}

We know from linear perturbation theory of the collapse of linear
overdensities in an FLRW universe, such as the Zel'dovich
approximation and $N$-body simulations, that matter generally
``falls'' from low density regions (voids) into filaments and streams
along filaments towards knots where the filaments join together into
what correspond today to massive galaxy clusters.

In general, the less massive objects move faster than the higher mass objects,
due to conservation of momentum (Newtonian):
 relatively low mass objects fall (in comoving coordinates) towards the massive
galaxy clusters at the ``knots'' of the cosmic web, while the most massive
clusters have relatively little peculiar velocities with respect to the 
comoving frame.

\fflatxLtwo

In this case, we can consider a massive cluster which is approximately 
stationary and $x$ to be the distance remaining between a test particle 
(of negligible mass) and the cluster.   The test particle is so far mostly 
comoving with the comoving reference frame, i.e. its peculiar velocity 
(velocity relative to the comoving frame), at which
it falls towards the cluster is ``small''.

The same assumptions and choices, (1) to (7), are made
as in \SSS\ref{s-cluster-self-gravity}. Since the cluster is considered
stationary, (8) is no longer relevant.

The geometry in this case is that shown in Fig.~\ref{f-flatxLtwo}.


\section{Calculation and results}
\label{s-results}

\subsection{One-dimensional analysis}

{
For simplicity, we first make a one-dimensional analysis, i.e.
we consider multiple images only in one direction, effectively
assuming a $\mathbb{T}^1 \times \mathbb{R}^2$ spatial hypersurface, 
hereafter, written $\mathbb{T}^1$.}

\subsubsection{Self-gravity of a large cluster}

Figure~\ref{f-flatxL} shows the geometry of the situation based on the
assumptions and general calculation choices of \SSS\ref{s-method}, for
a 1-torus, {$\mathbb{T}^1$} 
model, where, for simplicity, we assume that the
cluster is moving directly towards one of its adjacent images, rather
than at an arbitrary direction. A more accurate calculation would only 
modify the present calculation by less than an order of magnitude.

The Newtonian attractive force towards the slightly closer of the two
topological images of the adjacent cluster, i.e. towards the right
in Fig.~\ref{f-flatxL}, is then:

\begin{eqnarray}
F &\approx& 
            Gm^2 \left[ \frac{1}{(L-x)^2} - \frac{1}{(L+x)^2}  \right]
\label{e-firstimage}
\end{eqnarray}
where $m$ is the mass of the cluster, $G$ is the Newtonian gravitational constant,
$L$ is the comoving size of the fundamental domain in the chosen direction,
and $x << L$ is the displacement in the direction of the closest topological
image, in physical coordinate units.
By the time lapse assumption (7), the comoving distance $L$ is


Let us define 
\begin{equation}
\epsilon := \frac{x}{L} .
\label{e-defnepsilon}
\end{equation}

Then 
\begin{eqnarray}
F &\approx& G \frac{m^2}{L^2} \left[ \frac{1}{(1-\epsilon)^2} 
                                        - \frac{1}{(1+\epsilon)^2} \right] 
 \nonumber \\ 
 & \approx & G \frac{m^2}{L^2} 
           \left[ (1 + 2\epsilon + 3\epsilon^2 + \ldots) - 
                    (1 - 2\epsilon + 3\epsilon^2 + \ldots) \right] 
\nonumber \\ 
 & =  & G \frac{m^2}{L^2} 
          \left[ 4 \epsilon + \ldots \right] \nonumber \\
&\approx & 4G \frac{m^2 \epsilon}{L^2}. 
\label{e-firstimage-done}
\end{eqnarray}

If we rewrite this as an acceleration and substitute back the definition 
of $\epsilon$, 
then we have

\begin{eqnarray}
\ddot{x} & =&   \frac{4Gm}{L^3}\;x.  
\label{e-acceleqn}
\end{eqnarray}

The solution to this equation is the exponential:
\begin{equation}
{x} = e^{\sqrt{\frac{4Gm}{L^3}}t}.
\label{e-exponential}
\end{equation}

This is qualitatively what is expected from a cosmological constant: an
exponentially growing length scale. 

Could this have any relation to 
exponential growth in the scale factor $a(t)$, i.e. could it provide 
a cosmological constant, {or 
at least a form of dark energy consistent
with that observed?} 

Before discussing this question in the next section, we first note that, 
given the causal contact assumption (4), we could expect that not only the
closest topological image would have an effect on a cluster, but also 
successive images.

Eq.~(\ref{e-firstimage}) for the first $N$ successive pairs of topological 
images then becomes:
\begin{eqnarray}
F &\approx& Gm^2 \left[ \frac{1}{(L-x)^2} - \frac{1}{(L+x)^2}  \right]
\nonumber \\
            &&      + \frac{1}{(2L-x)^2} - \frac{1}{(2L+x)^2}  + \ldots  
\nonumber \\
&\approx&   \frac{4Gm^2}{L^2}\;\epsilon \; \sum^N_{i=1} \frac{1}{i^3} 
\nonumber \\
&\approx&   4.8 G \frac{m^2 \epsilon}{L^2} 
\label{e-manyimages}
\end{eqnarray}
for $N >> 1$.

This is only a small correction to Eq.~(\ref{e-firstimage-done}) 
 --- because an effect weakening with the 
cube of the distance decreases rapidly.

\subsubsection{Effect relative to the cosmic web}
\label{s-test-object-results}

As mentioned above, in \SSS\ref{s-test-object}, 
the geometry for an object slowly starting to fall towards a massive cluster, i.e. 
falling towards a dense node of the cosmic web of density perturbations, 
is that 
shown in Fig.~\ref{f-flatxLtwo}. 

The equation for the 
{\em long-distance component} of 
acceleration is algebraically the same as for cluster self-gravity, i.e.
as in Eq.~(\ref{e-firstimage}), except that there is also an 
acceleration term $-G\frac{m}{x^2}$ 
caused by the {\em local} copy of the cluster near the test object, since
we are interested in long-distance effects, and for convenience, 
we divide by the mass of the test object:
\begin{eqnarray}
\ddot{x} &\approx& -G\frac{m}{x^2} +
            Gm \left[ \frac{1}{(L-x)^2} - \frac{1}{(L+x)^2}  \right]
          \nonumber \\
&\approx &  -G\frac{m}{x^2} +  \frac{4Gm}{L^3}\;x.  
\label{e-firstimage-test-object}
\end{eqnarray}

The first term  in the second line of 
Eq.~(\ref{e-firstimage-test-object}) represents 
local attraction, i.e. Newtonian gravity as it is normally thought of, inversely
proportional to distance, but
the second term is a {\em long-distance term}, identical to 
Eq.~(\ref{e-acceleqn}), {\em directly} proportional to distance.

However, although this long-distance term is {\em
algebraically} identical to 
the right-hand side of Eq.~(\ref{e-acceleqn}), 
the interpretation is different.

Instead of the equation representing a high mass
cluster which has been perturbed from its position and is exponentially
accelerated away from its initial position, 
in this case we have a test particle which feels (in addition to local
acceleration towards the cluster potential well)
an acceleration {\em away} from 
its nearby (but multiply imaged) cluster potential well.

Again, {\em this is qualitatively similar to the effect of a
cosmological constant, since it is repulsive.} Moreover, it is an effect
{\em additional} to local gravitational terms --- which is what would be
required of something providing a cosmological constant.

{However, since we make assumptions about the
cosmological epoch, the effect is unlikely to be constant with time,
so it is more accurate to say that 
{\em the effect is qualitatively similar to the effect of a positive 
dark energy term}}.

Note that, in this case, if we consider
only the period of infall, before any path crossing or virialisation
occurs, i.e. when $x$ {\em decreases} with time, 
then force $F$ slows down the rate at which the test particle 
falls towards the cluster. Nevertheless, the particle is starting to fall 
towards the cluster --- in comoving coordinates --- so, if 
an individual test particle has started falling towards the cluster in 
{\em physical coordinates}, then the effect from the topological images 
will exponentially {\em decrease} with time,
as the particle approaches the cluster. This is not a problem, since
we are not interested in following the path of any individual particle
over time, and since we are most interested in a phenomenon which can be
related to the system of comoving coordinates itself, i.e. to the acceleration
equation of the FLRW model.

\fflatxLthD


\subsection{Three-dimensional analysis}
\label{s-threed}

We generalise the above calculation to an arbitrary displacement of
the test particle from the cluster, in a direction not necessarily
aligned with the generator $g_x$, and including Newtonian
approximation gravitational attraction from multiple images generated
by orthogonal generators $g_y$ and $g_z$, for the case of the
three-torus $\mathbb{T}^3$. We generalise the length $L$ of $g_x$ by writing
lengths $L_a \equiv L$, $L_e$, $L_u$ \nocite{LLL96}({Lehoucq} {et~al.} 1996) for the lengths of
$g_x$, $g_y$ and $g_z$ respectively. Define the displacement vector
\begin{equation}
{\bf r} := (x,y,z)
\end{equation}
of modulus $r$.

We continue to assume $x,y,z \ll L_a, L_e, L_u$ in order to make 
first order approximations from Taylor expansions.

The generalisation of Eq.~(\ref{e-firstimage-test-object}) is then
\begin{eqnarray}
\ddot{\bf{r}} 
&=& -G m \frac{\bf{r}}{r^3} + \ddotrtopo 
\nonumber \\
&=& -G m \frac{\bf{r}}{r^3} + 
\nonumber \\
&& \rule{-4ex}{0ex} Gm \rule{-3ex}{0ex} \sum_{(i,j,k)
           \not= 
           (0,0,0)}
  \frac{(iL_a-x,jL_e-y,kL_u-z)}{ 
        \left[ (iL_a-x)^2 + (jL_e-y)^2 + (kL_u-z)^2 \right]^{3/2}},
\nonumber \\
&&
\label{e-test-object-threed}
\end{eqnarray}
where we write $\ddotrtopo$ for the residual acceleration due to the topological
images of the cluster.

Since the forces add vectorially by orthogonal components, we can first
calculate the sum for the same two images we used before, i.e. for 
the two images of the cluster along the $x$-axis, 
at $i=\pm 1, j=k=0$. This is illustrated 
in Fig.~\ref{f-flatxLthD}.

\subsubsection{$x$ component for $i=\pm 1, j=k=0$}

The $x$ component of the two terms $i=\pm 1, j=k=0$ from
Eq.~\ref{e-test-object-threed} is, using $\epsilon := x/L_a$ as above, 


\begin{eqnarray}
&& \frac{1}{Gm} \left((\ddotrtopo)_{i=\pm1,j=k=0}\right)_x 
\nonumber \\
&=& 
\frac{L_a-x}{ 
        \left[ (L_a-x)^2 + y^2 + z^2 \right]^{3/2}}
+ \frac{-L_a-x}{ 
        \left[ (-L_a-x)^2 + y^2 + z^2 \right]^{3/2}}
          \nonumber \\
&=&
\frac{L_a^{-2}(1-\epsilon)}{ 
        \left[ (1-\epsilon)^2 + \left(\frac{y}{L_a}\right)^2 
                              + \left(\frac{z}{L_a}\right)^2 \right]^{3/2}}
          \nonumber \\
&&
+ \frac{L_a^{-2}(-1-\epsilon)}{ 
        \left[ (-1-\epsilon)^2 + \left(\frac{y}{L_a}\right)^2 
                              + \left(\frac{z}{L_a}\right)^2 \right]^{3/2}}
          \nonumber \\
&=& L_a^{-2} 
\left[ (1-\epsilon)
        \left(1-2\epsilon +...\right)^{-3/2} \right.
           \nonumber \\
&&
+ \left. (-1-\epsilon)
        \left(1+2\epsilon +...\right)^{-3/2} \right]
          \nonumber \\
&=& L_a^{-2} 
\left[ (1-\epsilon)(1+3\epsilon + ...) 
        +   (-1-\epsilon)(1- 3\epsilon + ...) \right]
          \nonumber \\
&=& L_a^{-2} 
\left[ (1+2\epsilon + ...) 
        +   (-1+ 2\epsilon + ...) \right]
          \nonumber \\
&=& 4\epsilon L_a^{-2} \nonumber \\
&=& 4 x L_a^{-3}.
\label{e-test-object-threed-x}
\end{eqnarray}

Unsurprisingly, to first order, this is the same result as in 
Eq.~(\ref{e-firstimage-test-object}). The $y$ and $z$ displacements do 
not affect the residual force in the $x$ direction, they only have
second order effects.

\subsubsection{$y$, $z$ components for $i=\pm 1, j=k=0$}

The $y$ and $z$ components of the $i=\pm 1, j=k=0$ terms are calculated
as follows.

\begin{eqnarray}
&&\frac{1}{Gm} \left((\ddotrtopo)_{i=\pm1,j=k=0}\right)_y 
\nonumber \\
&=& 
\frac{-y}{ 
        \left[ (L_a-x)^2 + y^2 + z^2 \right]^{3/2}}
+ \frac{-y}{ 
        \left[ (-L_a-x)^2 + y^2 + z^2 \right]^{3/2}}
          \nonumber \\
&=&
-y L_a^{-3} 
\left\{
        \left[ (1-\epsilon)^2 + \left(\frac{y}{L_a}\right)^2 
                              + \left(\frac{z}{L_a}\right)^2 \right]^{-3/2}
\right.
\nonumber \\
&&+ \left.
        \left[ (-1-\epsilon)^2 + \left(\frac{y}{L_a}\right)^2 
                              + \left(\frac{z}{L_a}\right)^2 \right]^{-3/2}
\right\}
          \nonumber \\
&=& -y L_a^{-3} 
\left[ 
        \left(1-2\epsilon +...\right)^{-3/2}
 +  
        \left(1+2\epsilon +...\right)^{-3/2} \right]
          \nonumber \\
&=& -y L_a^{-3} 
\left[ (1+3\epsilon + ...) 
        +  (1- 3\epsilon + ...) \right]
          \nonumber \\
&\approx & -2 y L_a^{-3},
\label{e-test-object-threed-y}
\end{eqnarray}
and similarly, 
\begin{eqnarray}
\frac{1}{Gm} \left((\ddotrtopo)_{i=\pm1,j=k=0}\right)_z
&\approx & -2 z L_a^{-3}.
\label{e-test-object-threed-z}
\end{eqnarray}

Unsurprisingly, this yields a weak net force pulling the test object back towards
the $x$-axis joining the two topological images of the cluster.

However, what may be surprising is that this force {\em increases} in 
amplitude as the test object's $y$ (or $z$) separation {\em increases}. For the 
$x$-component, it is obvious that as the test object becomes less and less
symmetrically placed between the two topological images, i.e. as $x$ increases,
the residual force towards the closer image should increase in amplitude.
But how is it possible that as the test object moves {\em further} from the
$x$-axis, i.e. as $y$ increases, the residual force pulling it back to
the plane increases?

The explanation is in the vectorial nature of the addition of
forces. (A Euclidean covering space is presumed throughout.) The force
in the $-y$ direction is only a component of a total, vectorial force.
For very small $y$, the two force vectors towards the two topological
images almost completely cancel since they are nearly perfectly
parallel. As $y$ increases, these two vectors
become less parallel and cancel less completely, so although their
individual (scalar) amplitudes decrease, the $y$ component of
their vector sum increases.

\subsubsection{Vectorial residual force for $x$ axis and for all images}
\label{s-residual-threed}

Eqs.~(\ref{e-test-object-threed-x}),
(\ref{e-test-object-threed-y})
and (\ref{e-test-object-threed-z})
yield
the total vectorial residual acceleration for the $i=\pm 1, j=k=0$ images of the
cluster, i.e. for the closest $x$-axis images:
\begin{eqnarray}
(\ddotrtopo)_{i=\pm1,j=k=0}  &=& 
Gm \frac{(4x,-2y,-2z)}{L_a^3}.
\label{e-test-object-threed-xyz}
\end{eqnarray}

By symmetry, the resultant residual acceleration for the closest
images from all three axes is:
\begin{eqnarray}
&&(\ddotrtopo)_{(i,j,k)\in\{(\pm1,0,0),(0,\pm1,0),(0,0,\pm1)\}}
\nonumber \\
&=& 
Gm
\left[
\frac{(4x,-2y,-2z)}{L_a^3} + 
\frac{(-2x,4y,-2z)}{L_e^3} + 
\right.
\nonumber \\
&& 
\rule{30ex}{0ex} \left. \frac{(-2x,-2y,4z)}{L_u^3}  
\right]
\nonumber \\
&=&
2 Gm \left[ 
   x \left( \frac{2}{L_a^3} -\frac{1}{L_e^3} - \frac{1}{L_u^3} \right), \;
   y \left( \frac{2}{L_e^3} -\frac{1}{L_a^3} - \frac{1}{L_u^3} \right),\;
\right.
\nonumber \\
&& 
\rule{25ex}{0ex} 
\left.  z \left( \frac{2}{L_u^3} -\frac{1}{L_a^3} - \frac{1}{L_e^3} 
   \right) \right].
\label{e-test-object-threed-unequal}
\end{eqnarray}

If $L_a=L_e=L_u$, these terms cancel and the resultant residual
acceleration is zero.

By symmetry, each successively 
distant orthogonal, equidistant 8-tuplet of topological images 
in Eq.~(\ref{e-test-object-threed}) also contributes a zero 
sum if $L_a=L_e=L_u$. Numerical calculation of the contributions of other
symmetrical $n$-tuplets of topological images distant from the $(0,0,0)$ image 
indicates that these also contribute zero to the sum, 
so that the full (first order) sum is zero.

This shows a three-dimensional effect different to that from the
$x$-axis calculation alone: the residual gravitational acceleration 
induced by multiple
images disappears if space is an ``isotropic $\mathbb{T}^3$'' model, in the sense
that the three lengths of the fundamental domain are equal.

On other hand, in a ``slightly anisotropic $\mathbb{T}^3$'' model, the
residual gravity due to multiple images does {\em not} totally disappear.
In order to consider the case in which the three side lengths of the
fundamental domain are slightly unequal, define
\begin{equation}
\delta_e := \frac{L_e}{L_a} - 1,  \;\;
\delta_u := \frac{L_u}{L_a} - 1.
\label{e-defndeltas}
\end{equation} 

Then we have 
\begin{eqnarray}
\frac{2}{L_a^3} -\frac{1}{L_e^3} - \frac{1}{L_u^3} 
&=& L_a^{-3} \left[ 2- (1+\delta_e)^{-3} - (1+\delta_u)^{-3} \right] 
\nonumber \\
&=& 
L_a^{-3} \left[ 2- (1 - 3\delta_e + ...) - (1-3\delta_u + ...) \right] 
\nonumber \\
&=& 
3 L_a^{-3} (\delta_e + \delta_u)
\end{eqnarray}
and similarly
\begin{eqnarray}
\frac{2}{L_e^3} -\frac{1}{L_a^3} - \frac{1}{L_u^3} 
&=& 
3 L_a^{-3} (-2 \delta_e + \delta_u),
\end{eqnarray}
so that
\begin{eqnarray}
&&(\ddotrtopo)_{(i,j,k)\in\{(\pm1,0,0),(0,\pm1,0),(0,0,\pm1)\}}
\nonumber \\
&=& 
6\, Gm L_a^{-3} \left[ 
   x \left(  \delta_e + \delta_u \right),
   y \left( -2 \delta_e + \delta_u \right),
   z \left( \delta_e -2 \delta_u \right) 
   \right].
\nonumber \\
\label{e-test-object-threed-unequal-two}
\end{eqnarray}

For the case $L_a=L_e$ (i.e. $\delta_e=0$), this becomes
\begin{eqnarray}
&&(\ddotrtopo)_{(i,j,k)\in\{(\pm1,0,0),(0,\pm1,0),(0,0,\pm1)\}}
\nonumber \\
&=& 
6\, Gm \delta_u L_a^{-3} \left( 
   x,  y, -2z \right) 
\nonumber \\
\label{e-test-object-threed-unequal-longz}
\end{eqnarray}
while for
the case $L_e=L_u$ (i.e. $\delta_e=\delta_u$), this becomes
\begin{eqnarray}
&&(\ddotrtopo)_{(i,j,k)\in\{(\pm1,0,0),(0,\pm1,0),(0,0,\pm1)\}}
\nonumber \\
&=& 
6\, Gm \delta_u L_a^{-3} \left( 
   2x,  -y, -z \right) 
\nonumber \\
\label{e-test-object-threed-unequal-shortx}
\end{eqnarray}

This differs from the acceleration in Eq.~(\ref{e-test-object-threed-xyz}) by
a factor of $3 \delta_u$. In other words, if the long fundamental dimensions 
are equal to one another and one-third greater than the short dimension, then
the linearised (in $\delta_u$) estimate of the effect summed 
from the three directions, in Eq.~(\ref{e-test-object-threed-unequal-shortx}),
is as large as if the two larger dimensions were infinite, as is effectively
represented by Eq.~(\ref{e-test-object-threed-xyz}). Clearly, this 
implies that the approximation is valid only for $\delta_u \ll 1/3$.

In each case, this qualitative behaviour is similar to a positive 
dark energy term in the direction of the shorter fundamental length(s), 
and a negative dark energy term in the direction of the longer
fundamental length(s), indicating that {\em the effect would tend to
equalise the three fundamental lengths of a $\mathbb{T}^3$ model}. 

Numerical 
\bfrevisedversion{evaluation of further terms in the sum on the right-hand
side of} Eq.~(\ref{e-test-object-threed}) indicates that these modify 
the total sum only slightly.



\section{Discussion}
\label{s-discuss}

Is there any relation between the effect found here 
and {dark energy}? 
{Would 
this effect really tend to
equalise the three fundamental lengths of a $\mathbb{T}^3$ model?}

{
For simplicity, consider the $x$-axis case and the cluster or test object
displaced along the $x$-axis, for the $\mathbb{T}^1$ model.
}

What has been shown so far is
that, under the assumptions listed
above, the gravitational effect
due to multiple topological imaging
provides an acceleration proportional to displacement, 
i.e. constant $\ddot{x}/x$, where 
either (\SSS\ref{s-cluster-self-gravity}) this is 
the total acceleration 
for a given massive object towards its closer topological image, 
or, (\SSS\ref{s-test-object}) it is 
the total long-distance induced acceleration
(in addition to locally induced acceleration) for a test object ``falling''
towards a given massive object.

The calculation itself is made in comoving space: the result 
of \SSS\ref{s-cluster-self-gravity}, for
self-gravity of a cluster towards itself, is that a {\em perturbed},
large massive object is slightly accelerated in the direction of its
closest image, and this acceleration is proportional to the
displacement from the initial position in comoving space, so that the
displacement increases exponentially.  Since we are working within the
comoving frame, it is not obvious, in the Newtonian approximation, how
to relate this to a modification in the equations for the growth of
the scale factor itself with cosmological time.

On the other hand, since the cosmic web is, on average, fixed in the
comoving frame, it may be possible to interpret 
the second case (\SSS\ref{s-test-object}), 
of test objects ``falling'' towards the
most massive objects in the cosmic web, 
in terms of {a dark energy term}.

For objects still distant from and falling into massive clusters,
could the additional force term
of Eq.~(\ref{e-acceleqn}), as shown in Fig.~\ref{f-flatxLtwo}, 
provide the pressure term in the FLRW acceleration equation, 
\begin{equation}
\frac{\ddot{a}}{a} = - \frac{4\pi G}{3} \rho 
    \left( 1 + 3 \frac{p}{\rho c^2} \right),
\label{e-flrw-accel}
\end{equation}
where $a$ is the scale factor, in order to mimic {dark energy}?

\subsection{Amplitude of cosmo-topological gravity relative to the cosmic web}

{An heuristic}
Newtonian derivation of the acceleration equation in this case,
{for simplicity, treating the $x$-axis case, for the 
$\mathbb{T}^1$ model},
 follows
from Eq.~(\ref{e-firstimage-test-object}), 
where we define $\chi$ to be a fixed length in the comoving reference frame 
so that $x = a \chi$.  
$L$ has already been defined to be a comoving length, but in the above equations
we implicitly used $a=1$; here, since we want expressions valid at arbitrary 
values of the scale factor, not only $a=1$,
we write $aL$ rather than $L$.
\begin{eqnarray}
\ddot{a} \chi  
 &=& \ddot{x}  \nonumber \\
 &\approx& -G \frac{m}{x^2} +  G \frac{4m}{(aL)^3}\;x \nonumber \\
 &=&  -G\frac{4\pi {\rho}}{3} \; a \chi 
       + G\frac{4\pi {\rho}}{3} \; \frac{4a^3\chi^3}{a^3L^3}\;a \chi  
\label{e-accelneqn-b}
\end{eqnarray}
if we estimate that the cluster mass was obtained from matter spread 
at the mean density $\bar{\rho}$ throughout a sphere of radius $x$.

Note that assumptions (5) and (6) are crucial here, since we assume
that the relevant matter density, both locally and at long distance,
is that contained inside of local and topologically imaged, distant
spheres, fixed within comoving coordinates, around the cluster centre.

Dividing both sides of Eq.~(\ref{e-accelneqn-b}) by $x= a\chi$ yields
\begin{eqnarray}
\frac{\ddot{a}}{a} 
 &= & -G\frac{4\pi {\rho}}{3} 
            \left( 1 - \frac{4\chi^3}{L^3} \right).  \\
\label{e-acceleqn-topo}
\end{eqnarray}

Using the standard notation for a dark energy component, 
$w := p/(\rho c^2)$,
we can rewrite this
\begin{equation}
 w = - \frac{4}{3}  \left(\frac{\chi}{L}\right)^3 
   \sim - \left(\frac{\chi}{L}\right)^3.
\label{e-w-estimate}
\end{equation}

Since we are interested in objects ``falling'' (in comoving coordinates) towards
the nodes of the cosmic web, i.e. at most a few tens of Mpc from those nodes,
then for a universe side length
as large as the diameter of the surface of last scattering, $L = 20${\hGpc}, 
we have 
\begin{equation}
w \sim - 
 10^{-9}.
\label{e-order_magnitude}
\end{equation}

So, 
{for the $x$-axis case in a $\mathbb{T}^1$ model,}
while cosmo-topological gravity has the right {\em algebraic}
characteristics 
{in this heuristic approach}, 
{\em 
{with} the assumptions listed above}, to provide
{an acceleration similar to a dark energy term},
its {\em amplitude in the present-day
  Universe} is certainly too small to be significant, except possibly
for extremely high-resolution $N$-body simulations of the formation of
structure in the Universe.

Of course, if the {length} 
scale of large scale structure were nearly as large
as that of the Universe itself, i.e. if $\chi \sim L$, then the amplitude
of this effect would be much larger. However, this would 
not be physically realistic
according to our understanding of structure formation.

{As shown in \SSS\ref{s-threed}, if we consider a 
displacement of the test particle in an arbitrary direction relative to 
the generator vector,
and if we sum the contributions from the three directions in a
$\mathbb{T}^3$ model of nearly equal fundamental lengths 
in the three directions, then a similar heuristic 
argument, using an anistropic scale factor 
\begin{equation}
{\bf a}(t) = (a_a, a_e, a_u)(t)
\label{e-a-anisotropic}
\end{equation}
implies a similar, but anisotropic, effective acceleration term
approximately $\delta$ times weaker than for a $\mathbb{T}^1$ model
[Eqs~(\ref{e-firstimage-test-object}),
(\ref{e-test-object-threed}),
(\ref{e-test-object-threed-unequal-shortx})],
where $\delta$
is the fractional difference in fundamental lengths.}

{In contrast to the $\mathbb{T}^1$ case, if the three lengths 
are exactly equal, then this effect cancels to zero. The fundamental lengths
must be slightly unequal in order for there to be an effect.}

{
Interestingly, as noted in
\SSS\ref{s-residual-threed}, this effect 
will be anisotropic in such a way as to
oppose the anisotropy of the three fundamental lengths, tending to
push the three fundamental lengths towards equality. This is probably 
the first time that such an effect tending to induce $L_a=L_e=L_u$ for 
a $\mathbb{T}^3$ model has been found. The latter is often {\em assumed} for simplicity
and aesthetic reasons, but here we seem to have a physical motivation for
this equality as a stable equilibrium rather than as an arbitrary assumption.}

\subsection{Caveat: Time-varying mass of cluster}

A minor caveat to note for cosmo-topological gravity 
is that the mass $m$ is
not constant with time in our model: gravitational collapse will continue and
successively larger and larger objects will form. However, 
{it has 
already been noted above that this effect is linked to the cosmological
epoch, and is thus likely to vary with time,}
so this is not
necessarily a strong argument against some role for this effect 
as a dark energy term: the low amplitude of the effect is
a much greater problem.

\ftwotorusmod

\ftwotorusx

\fthreetorusmod

\fthreetorusx

\subsection{Caveat: $\chi$ not constant {within} 
large scale structure unit}

Another minor caveat,
{in addition to the anisotropy of the effect},
is that this effect will vary with 
distance $\chi$ from the nearest big cluster (node of the cosmic web).

\recenteditsone{
This implies that the {\em spatially averaged} value of 
$w \sim - \left(\frac{\chi}{L}\right)^3$ would be needed before comparing
a theoretical $w$ value with an observed value, such as the 
presently estimated value of $w \approx -1$. 
}

Since most cosmological observations relevant to estimating the parameters
of the metric are in practice averaged out over scales larger than that of
large scale structure, i.e. on scales $\gg 100$~{\hMpc}, this is not a problem.

\bfrevisedversion{However, it is interesting to see how the ratio of 
the topological acceleration term to the local acceleration term towards
the dominant nearby cluster varies as a function of the length scale towards
nearby clusters. This has been presented below in \SSS\ref{s-numerical-method}.}

\bfrevisedversionstart

\subsection{Numerical test}  \label{s-numerical-method}

As a numerical check on analytical calculations of this effect, 
the zero redshift output data file for a ``Lambda CDM'' model of 
the ``Hubble Volume'' one billion particle 
simulation of the present-day distribution of galaxy clusters made
by the Virgo Supercomputing Consortium \nocite{HubbleVol02}({Evrard} {et~al.} 2002) is used here.

In other words, a present-day synthetic distribution of clusters for a
model universe with $\Omm=0.3, \Omega_\Lambda = 0.7, H_0= 70$~km/s/Mpc and a
CDM (cold dark matter) initial power spectrum of density perturbations
consistent with these parameters, is used.\footnote{See:
\url{http://www.mpa-garching.mpg.de/Virgo/hubble.html}; 
the file is downloadable at the time of submission as 
\url{http://ln-s.net/E3s} or as 
{http://www.mpa-garching.mpg.de/Virgo\_Data/\-hubble\_cluster/snapshot/lcdm/\-cluster\_fof\_lcdm\_z0\_b0.164.tar.gz}.} 
The data file contains mass and positions of 
``clusters'' detected as virialised groups of particles by
using a friends-of-friends (FOF) algorithm with a linking length of
$b=0.164$.  Individual particles have masses of $2.24 \times
10^{12}h^{-1}M_{\odot}$.




The side length of the cube is 3$h^{-1}$Gpc.  This length scale is
about six times smaller than the likely minimum value of $L$, but is 
about one and a half orders of magnitude larger than the length scale
of ``large scale structure'', i.e. the scales on which clusters form voids
and walls, so it should be good enough to qualitatively test 
the result of analytical calculations.

%
%
%
%

For each test point, the vector accelerations as described in 
Eq.~(\ref{e-test-object-threed}) are calculated.
That is, the standard gravitational acceleration
${-G m \frac{\bf{r}}{r^3}}$ towards the cluster whose gravitational pull is
strongest at the test point and the ``topological'' gravitational 
component $\ddotrtopo$ for the topological images of this cluster 
are calculated. This is done:
\begin{list}{(\roman{enumi})}{\usecounter{enumi}}
\item for the $\mathbb{T}^1$ case (2 adjacent topological images), 
\item for the $\mathbb{T}^3$ case (8 adjacent topological images) 
with three exactly equal fundamental lengths, and 
\item for the $\mathbb{T}^3$ case with $y$ and $z$ axes' fundamental lengths
a small fraction $\delta_e = \delta_u = 0.1$ larger than the $x$ axis fundamental
length.
\end{list}

From Eqs~(\ref{e-test-object-threed-xyz}) and 
(\ref{e-test-object-threed-unequal-shortx}) and the derivation leading to 
Eq.~(\ref{e-w-estimate}), we would expect these three ratios (for the $x$ axis
direction which dominates, without loss of generality) to be 
approximately:
\begin{list}{(\roman{enumi})}{\usecounter{enumi}}
\item $4 \left(\frac{\chi}{L}\right)^3$,
\item $\ll 4 \left(\frac{\chi}{L}\right)^3$, and
\item $12 \delta_e \left(\frac{\chi}{L}\right)^3$
\end{list}
respectively.

Figures \ref{f-twotorusmod}-\ref{f-threetorusx} show that in a realistic
simulation on a scale just one order of magnitude smaller than 
the diameter of the last scattering surface \nocite{HubbleVol02}({Evrard} {et~al.} 2002), and on 
scales far enough from individual clusters to be participating in the
linear regime of density perturbation theory (i.e. still expanding with
the Hubble flow), the ratio of the ``topological'' acceleration to the
normal local acceleration is consistent with these three expressions.

Of course, since the size of the simulation is 3{\hGpc} rather than a
more realistic scale of 20{\hGpc}, the acceleration ratios in these
figures should be reduced by approximately $(20/3)^3$, i.e. by a
factor of about 300, in agreement with the estimate presented below in
Eq.~(\ref{e-order_magnitude}).

\bfrevisedversionstop

\subsection{Assumptions (5), (6), the topological one-body and two-body problems}
\label{s-onebody}

Another caveat, probably more important, is that the calculation we
have made, based on assumptions (5) and (6)
(\SSS\ref{s-cluster-self-gravity}), is equivalent to what we might
call the ``topological one-body and two-body problems''.

%

\bfrevisedversionstart 

Na\"{\i}vely, it may seem that both the Newtonian and relativistic
versions of gravity exclude gravitational self-interaction of a
single, massive, point-sized body: a gravitational interaction between $N$
bodies normally requires $N \ge 2$ bodies.

However, this intuition relies on the implicit assumption that space is
an infinite, simply connected Euclidean space.

In a multiply connected space, a single body can be thought of --- in the
{\em covering space} --- as a set of multiple bodies. Although $N=1$ 
bodys exists in a true physical sense and in the fundamental domain, $N \gg 1$ 
bodies exist in the covering space, which is the simplest space in which to 
calculate gravitational interactions. 

This is why it is possible for a single body to have gravitational
self-interactions according to the standard Newtonian approximation of
gravity. Hence, the ``one-body problem'' is a serious dynamical problem,
despite its apparent absurdity.

The method of calculating these interactions depends on how realistic
our universe model is.

\bfrevisedversionstop

The assumptions (5) and (6) permit a relatively simple calculation, 
and are {\em equivalent} to assuming a flat (even though mostly empty) space 
which contains just one (or two) objects, i.e. our assumptions are equivalent
to a very simple, though physically unrealistic, model.

Also, the distant, supra-SLS images of a cluster are not-yet collapsed 
perturbations, so approximating them as single, point-like objects, while
ignoring matter outside the radius from which they form is somewhat 
arbitrary.

Is it possible to make a more precise calculation which avoids assumptions (5) 
and (6)? Numerical calculations would be possible in principle, and would provide
a good followup to our present result. 

However, these are unlikely to be easy, and would be different from
standard $N$-body simulations because of assumptions (7) and (8). Standard
$N$-body simulations use various numerical approximation techniques in order
to make the calculation time short enough to be practical, and, {\em de facto} 
make at least one of the following two asssumptions:
\begin{list}{(\roman{enumi})}{\usecounter{enumi}}
\item instead of assumption (4), assume that something like inflation has {\em not}
occurred
\item instead of assumptions (5) and (6), assume that all 
contributions to the local gravitational potential due to
long-range, supra-SLS distance scale density perturbations {\em perfectly 
cancel out}.
\end{list}

Since inflation scenarios are good candidates for providing some
ingredient of the correct model of the Universe, there is a good chance that
assumption (i) is incorrect.

Moreover, it is not clear to what extent assumption (ii) is a good
approximation. 

Newtonian gravititational attraction between two point objects
decreases according to the inverse square of the distance, i.e. as 
$\propto r^{-2}$, but the mass in successive spherical shells of equal
thickness increases as $r^2$: an anisotropy of fixed (small) solid 
angular size $\cos\theta$d$\phi\,$d$\theta$ 
at all radii would provide equal attractive forces
from each successive shell, no matter how distant. 

This is no problem in a perfectly homogeneous universe: hence, the FLRW solution
of the Einstein field equations.

But in a universe with perturbations, i.e. a ``slightly''
inhomogeneous, anisotropic universe, the contributions do not
perfectly cancel.

In a nearly FLRW universe, i.e. one containing a spectrum of density
perturbations, the degree to which the long-distance terms cancel 
most likely depends on the full nature of the perturbation spectrum.

\nocite{Melott90}{Farrar} \& {Melott} (1990) suggest from their $N$-body calculations that
assumption (ii) is correct in the case of a multiply connected flat
space (of equal fundamental lengths). Further work in this direction
would be interesting, in particular by introducing the physically {\em
standard} assumption (not used explicitly in normal $N$-body
simulations), i.e. our assumption (7), that $\cGW$ is finite and of the
standard value $\cGW =\cST=\cGR=\cEM$. Testing the effects of slightly
unequal fundamental lengths would also be interesting.

\subsubsection{Inhomogeneities and the Einstein-Hilbert equations}
As has been pointed out for many years, e.g. 
\nocite{Ellis99Carg}Section~7,~ {Ellis} \& {van Elst} (1999) and references therein, 
finding an exact inhomogeneous solution to the
Einstein-Hilbert equations, and then averaging this to find a ``mean density''
homogeneous solution,  is {\em not} mathematically
equivalent to adding
perturbations to the exact FLRW (homogeneous)
solution to the Einstein-Hilbert equations. The latter, approximate approach gives
a good match to observations, but this does not prove that the difference 
between the exact and perturbative approaches is negligible.

\nocite{BuchCarf02,BuchCarf03}{Buchert} \& {Carfora} (2002, 2003) have been studying this problem
in depth for several years
and have noted that the difference between the perturbative approach
and the exact (but still developing) approach leads to a ``backreaction'' term
(and also a curvature term)
which may have an effect on scales ranging from 
that of a massive galaxy cluster 
possibly up to super-Hubble length scales. 
Recent discussions \nocite{Kolb05a,Kolb05b}({Kolb} {et~al.} 2005b, 2005a) indicate that this presently 
seems, pending full calculation, to be a viable candidate 
for a dark energy term, without the need for, e.g. adding any scalar fields, based
on the backreaction term from {\em sub-}Hubble length perturbations.

In a qualitative sense, this is similar to the effect discussed in
this paper, except that we are primarily interested here in the effect of 
perturbations on gravity in the case of non-trivial global topology.

\nocite{BuchertEhlers97}{Buchert} \& {Ehlers} (1997) and \nocite{EhlersBuchert97}{Ehlers} \& {Buchert} (1997) presented work
closely related to what is presented here, considering both
$\mathbb{T}^3$ and $\mathbb{R}^3$ models, for the Newtonian case of
perfectly isotropic expansion:
\begin{equation}
a_a(t) = a_e(t) = a_u(t) \;\; \forall t.
\label{e-isotropic-a}
\end{equation}

For the $\mathbb{T}^3$ case, they found a consistent result with our 
first order result, i.e. that for a compact space admitting a global Hubble
flow, i.e.  if the average shear and rotation are zero (Appendix~A, 
\nocite{BuchertEhlers97}{Buchert} \& {Ehlers} 1997; see also four equivalent statements related to 
this in \SSS3.4 of \nocite{BuchertEhlers97}{Buchert} \& {Ehlers} 1997), then the perturbations
have no effect on the rate of expansion according in the Newtonian 
approximation.

Given that our heuristic calculation suggests that for a slight anisotropy
in the three lengths scales of  $\mathbb{T}^3$, the expansion rate should
be anisotropic, it would be interesting to develop \nocite{BuchertEhlers97}{Buchert} \& {Ehlers} (1997)'s 
approach further using (slightly) anisotropic scale factors, i.e. 
using Eq.~(\ref{e-a-anisotropic}) to replace
\nocite{BuchertEhlers97}{Buchert} \& {Ehlers} (1997)'s equation $V(t)=:a_{\cal D}^3 (t)$ (just after eq.~(3)) 
by
\begin{equation}
V(t)= a_aa_ea_u(t)
\label{e-anisotropic-vol}
\end{equation}
and following through, using the definition in Eq.~(\ref{e-defndeltas}).
The related work in \nocite{EhlersBuchert97}{Ehlers} \& {Buchert} (1997) could be reworked
starting from their eq.~(15).

Just as perfect homogeneity is certainly wrong, perfect isotropy in
the expansion rate is probably also just an approximation to a more
accurate model, whether or not the correct model is closer to
$\mathbb{T}^3$ or $\mathbb{R}^3$ or another model.

\section{Conclusion}
\label{s-conclu}

A residual gravitational effect, which we could possibly 
call ``cosmo-topological gravity'',
occurs due to distant multiple topological images in a multiply connected
universe which ``remembers'' the gravitational potential generated by 
multiple topological images, which in the covering (apparent) space are 
located outside of the present surface of last scattering, and were 
causally contacted at some earlier epoch, for example due to some prior
amount of inflation.
A Newtonian
approximation, in which the speed of transmission of gravitational information
is finite (equal to the special relativistic space-time constant $c$),
rather than infinite as in a fully Newtonian calculation,
is used here.

For a low mass test object ``falling'' (in comoving coordinates) 
towards a relatively nearby (at a comoving distance $\chi$ 
less than a few tens of Mpc) large,
massive collapsed object at the present epoch, i.e. a massive cluster
of galaxies, in a 3-torus universe of side-length $L \approx 20$~{\hGpc},
the two closest topological images of the cluster together yield a
residual Newtonian force on the test particle which locally appears as a force 
{\em repelling} the test particle away from (its nearby image of) the cluster,
{provided that we consider the test object lying 
along the axis joining the two closest topological images of the cluster
in one of the fundamental directions, and we only consider these two 
images.}

This residual force provides an acceleration algebraically
similar to that of {dark energy}, but weaker by
many orders of magnitude, 
i.e. by a factor of approximately $\left(\frac{\chi}{L}\right)^3 \sim 10^{-9}$
{at the present epoch.}

{A more general, three-dimensional calculation, for 
a test object displaced in an arbitrary direction in a 
3-torus universe, shows that the
effect cancels out to zero if the three side lengths are exactly equal.
If the side lengths are slightly unequal, by a fraction $\sim\delta$, then an 
{\em anisotropic} dark energy term, about
$\delta \left(\frac{\chi}{L}\right)^3 \sim 10^{-9}$ times weaker than
the observed dark energy 
(for side length $L \approx 20$~{\hGpc}), 
will accelerate the expansion of the shorter
length(s) and decelerate the expansion of the longer length(s), tending
to equalise them.}

{This is probably the first known physical effect 
which could relate the three fundamental lengths of a 
universe with $\mathbb{T}^3$ spatial sections.
The equality of the three lengths in simulations of the $\mathbb{T}^3$ model
has often been assumed, but on purely aesthetic grounds, without
any physical justification.
}


\bfrevisedversion{It} 
is clear that the effect is not significant in the present-day
Universe.


Apart from developing an anisotropic version of \nocite{BuchertEhlers97}{Buchert} \& {Ehlers} (1997)'s 
approach, an interesting (and challenging) followup 
project would be to check whether or not the same effect occurs in 
spherical, multiply connected universes: could the residual gravitational 
force due to multiple imaging have helped push the shape of the Universe
into that of the Poincar\'e dodecahedral space (PDS) during or not long
after the quantum epoch? Did it help isotropise the Universe?

\section*{Acknowledgments}


SB acknowledges support from KBN Grant 1P03D 012 26.  The simulations
used in this paper were carried out by the Virgo Supercomputing
Consortium using computers based at the Computing Centre of the
Max-Planck Society in Garching and at the Edinburgh parallel Computing
Centre. The data are publicly available at
\url{http://www.mpa-garching.mpg.de/NumCos}.
\bfrevisedversion{Helpful comments from an anonymous referee
were greatly appreciated.} 

\subm{\clearpage}

{
%

}


\begin{thebibliography}{}

\bibitem[{Ahmadi} \& {Nouri-Zonoz} 2005]{Nouri05}
{Ahmadi}, N., \& {Nouri-Zonoz}, M. 2005, arXiv:gr-qc/0510100

\bibitem[{Aurich}, {Lustig}, \&  {Steiner} 2005a]{Aurich2005a}
{Aurich}, R., {Lustig}, S., \& {Steiner}, F. 2005a, \cqg, 22,  3443, \eprint{astro-ph/0504656}

\bibitem[{Aurich}, {Lustig}, \&  {Steiner} 2005b]{Aurich2005b}
{Aurich}, R., {Lustig}, S., \& {Steiner}, F. 2005b, \cqg, 22,  2061, \eprint{astro-ph/0412569}

\bibitem[{Bagla} 2005]{Bagla05}
{Bagla}, J.~S. 2005, Current Science, 88, 1088

\bibitem[{Barrow} \& {Levin} 2001]{LevinTwins01}
{Barrow}, J.~D., \& {Levin}, J. 2001, \pra, 63, 044104, \eprint{gr-qc/0101014}

\bibitem[{Blanl{\oe}il} \& {Roukema} 2000]{BR99}
{Blanl{\oe}il}, V., \& {Roukema}, B.~F., eds. 2000, ``Cosmological Topology in  Paris 1998'' (Paris: Blanl{\oe}il \& Roukema), \eprint{astro-ph/0010170}

\bibitem[{Buchert} \& {Carfora} 2002]{BuchCarf02}
{Buchert}, T., \& {Carfora}, M. 2002, \cqg, 19, 6109, \eprint{gr-qc/0210037}

\bibitem[{Buchert} \& {Carfora} 2003]{BuchCarf03}
{Buchert}, T., \& {Carfora}, M. 2003, Physical Review Letters, 90, 031101,  \eprint{gr-qc/0210045}

\bibitem[{Buchert} \& {Ehlers} 1997]{BuchertEhlers97}
{Buchert}, T., \& {Ehlers}, J. 1997, \aap, 320, 1

\bibitem[{Ehlers} \& {Buchert} 1997]{EhlersBuchert97}
{Ehlers}, J., \& {Buchert}, T. 1997, General Relativity and Gravitation, 29,  733, \eprint{astro-ph/9609036}

\bibitem[{Ellis} \& {Uzan} 2005]{EllisUzanC03}
{Ellis}, G.~F.~R., \& {Uzan}, J.-P. 2005, Am.J.Phys., 73, 240,  \eprint{gr-qc/0305099}

\bibitem[{Ellis} \& {van Elst} 1999]{Ellis99Carg}
{Ellis}, G.~F.~R., \& {van Elst}, H. 1999, in NATO ASIC Proc. 541: Theoretical  and Observational Cosmology, ed. M.~{Lachi{\`e}ze-Rey}, 1--116,  \eprint{arXiv:gr-qc/9812046}

\bibitem[{Evrard}, {MacFarland}, {Couchman}, {Colberg},  {Yoshida}, {White}, {Jenkins}, {Frenk}, {Pearce}, {Peacock}, \&  {Thomas} 2002]{HubbleVol02}
{Evrard}, A.~E., {MacFarland}, T.~J., {Couchman}, H.~M.~P., {et al.} 2002, \apj, 573, 7,  \eprint{astro-ph/0110246}

\bibitem[{Farrar} \& {Melott} 1990]{Melott90}
{Farrar}, K.~A., \& {Melott}, A.~L. 1990, Computers in Physics, 4, 185

\bibitem[{Gundermann} 2005]{Gundermann2005}
{Gundermann}, J. 2005, e-print, \eprint{astro-ph/0503014}

\bibitem[{Kolb}, {Matarrese}, \&  {Riotto} 2005a]{Kolb05b}
{Kolb}, E.~W., {Matarrese}, S., \& {Riotto}, A. 2005a, ArXiv  Astrophysics e-prints, \eprint{astro-ph/0511073}

\bibitem[{Kolb}, {Matarrese}, \&  {Riotto} 2005b]{Kolb05a}
{Kolb}, E.~W., {Matarrese}, S., \& {Riotto}, A. 2005b, ArXiv  Astrophysics e-prints, \eprint{astro-ph/0506534}

\bibitem[{Lachi{\`e}ze-Rey} 1999]{LachiezeReyCTP99}
{Lachi{\`e}ze-Rey}, M. 1999, in Cosmological Topology in Paris 1998, 14  December 1998, Observatoire de Paris, Eds.: V. Blanl{\oe}il, B.F. Roukema,  \eprint{astro-ph/0010170}

\bibitem[{Lachi\`eze-Rey} \& {Luminet} 1995]{LaLu95}
{Lachi\`eze-Rey}, M., \& {Luminet}, J. 1995, \physrep, 254, 135,  \eprint{gr-qc/9605010}

\bibitem[{Lahav} \& {Liddle} 2004]{LahavLiddle04}
{Lahav}, O., \& {Liddle}, A. 2004, Physics Letters B, 592, 1,  \eprint{astro-ph/0406681}

\bibitem[{Lehoucq}, {Lachi\`eze-Rey}, \&  {Luminet} 1996]{LLL96}
{Lehoucq}, R., {Lachi\`eze-Rey}, M., \& {Luminet}, J.-P. 1996, \aap, 313, 339,  \eprint{gr-qc/9604050}

\bibitem[{Linde} 2004]{LindeTopo04}
{Linde}, A. 2004, Journal of Cosmology and Astro-Particle Physics, 10, 4,  \eprint{hep-th/0408164}

\bibitem[{Luminet} \& {Roukema} 1999]{LR99}
{Luminet}, J., \& {Roukema}, B.~F. 1999, in NATO ASIC Proc. 541: Theoretical  and Observational Cosmology, 117, \eprint{astro-ph/9901364}

\bibitem[{Luminet}, {Weeks}, {Riazuelo}, {Lehoucq}, \&  {Uzan} 2003]{LumNat03}
{Luminet}, J., {Weeks}, J.~R., {Riazuelo}, A., {Lehoucq}, R., \& {Uzan}, J.  2003, \nat, 425, 593, \eprint{astro-ph/0310253}

\bibitem[{Luminet} 1998]{Lum98}
{Luminet}, J.-P. 1998, Acta Cosmologica, XXIV-1, 105, \eprint{gr-qc/9804006}

\bibitem[{Rebou\c{c}as} \& {Gomero} 2004]{RG04}
{Rebou\c{c}as}, M.~J., \& {Gomero}, G.~I. 2004, Braz. J. Phys., 34, 1358,  \eprint{astro-ph/0402324}

\bibitem[{Roukema} 2000]{Rouk00BASI}
{Roukema}, B.~F. 2000, \BASI, 28, 483, \eprint{astro-ph/0010185}

\bibitem[{Roukema} 2002]{Rouk02topclass}
{Roukema}, B.~F. 2002, in {Marcel Grossmann IX Conference on General  Relativity}, eds V.G. Gurzadyan, R.T. Jantzen and R. Ruffini, World  Scientific, Singapore, p. 1937, \eprint{astro-ph/0010189}

\bibitem[{Roukema}, {Lew}, {Cechowska}, {Marecki}, \&  {Bajtlik} 2004]{RLCMB04}
{Roukema}, B.~F., {Lew}, B., {Cechowska}, M., {Marecki}, A., \& {Bajtlik}, S.  2004, \aap, 423, 821, \eprint{astro-ph/0402608}

\bibitem[{Roukema}, {Peterson}, {Quinn}, \&  {Rocca-Volmerange} 1997]{RPQR97}
{Roukema}, B.~F., {Peterson}, B.~A., {Quinn}, P.~J., \& {Rocca-Volmerange}, B.  1997, \mnras, 292, 835, \eprint{astro-ph/9707294}

\bibitem[{Starkman} 1998]{Stark98}
{Starkman}, G.~D. 1998, \cqg, 15, 2529

\bibitem[{Uzan}, {Lehoucq}, \& {Luminet} 1999]{ULL99b}
{Uzan}, J.-P., {Lehoucq}, R., \& {Luminet}, J.-P. 1999, in "Proc. of the  XIX$^{\rm th}$ Texas meeting, Paris 14--18 December 1998, Eds. E. Aubourg, T.  Montmerle, J. Paul and P. Peter, article n$^{\rm o}$ 04/25",  \eprint{gr-qc/0005128}

\bibitem[{Uzan}, {Luminet}, {Lehoucq}, \&  {Peter} 2002]{UzanTwins02}
{Uzan}, J.-P., {Luminet}, J.-P., {Lehoucq}, R., \& {Peter}, P. 2002, Eur. J.  Phys., 23, 277, \eprint{physics/0006039}

\end{thebibliography}
\end{document}